\documentclass[trackchanges]{aastex7}
\usepackage{siunitx}      
\usepackage{amsmath}

\usepackage{booktabs}
\usepackage{graphicx}
\usepackage{subcaption}
\usepackage{tabularray}
\begin{document}

\title{Revealing Exotic Nanophase Iron in Lunar Samples Through Impact-Driven Spatial Fingerprints}
 
\author[orcid=0000-0002-8624-1264,sname='North America']{Ziyu Huang}
\affiliation{Georgia Institute of Technology, Atlanta, Georgia 30332, USA}
\email[show]{zyuhuang@gatech.edu}  

\author[orcid=0000-0002-1821-5689,sname='North America']{Masatoshi Hirabayashi}
\affiliation{Georgia Institute of Technology, Atlanta, Georgia 30332, USA}
\email[]{}

\begin{abstract}
 
Nanophase iron (npFe) plays a crucial role in controlling the optical, chemical, and physical evolution of lunar regolith grains. While in-situ formation of npFe via reduction of native Fe-bearing minerals has long been considered a dominant pathway, recent mineralogical evidence from \cite{zeng2025exotic} reveals that 
the source of
a significant fraction of npFe may be delivered directly by exotic micrometeoroid impacts
(exotic npFe).
Yet the atomic-scale processes governing how exotic np-Fe forms and survives during hypervelocity impacts remain largely unknown.
To quantitatively compare in-situ and exotic delivery and formation of npFe, we perform a series of innovative atomistic modeling of micrometeoroid impacts with distinct projectile–target compositions: (1) SiO$_2$ projectiles on Fe$_2$SiO$_4$ targets (in-situ formation), (2) Fe$_2$SiO$_4$ projectiles on SiO$_2$ targets (exotic delivery).
Our results reveal distinct mechanistic fingerprints: in-situ np-Fe forms diffusely and radially around the impact site, whereas exotic np-Fe is efficiently retained and concentrated in asymmetric, momentum-aligned clusters. These contrasting spatial signatures provide a potential diagnostic criterion for distinguishing exotic versus in-situ np-Fe in returned lunar soils. In agreement with Chang’e-5 observations, our simulations demonstrate that exotic np-Fe production can be substantial, particularly in Fe-poor terrains such as highland regions. These findings highlight the need to account for exotic np-Fe when interpreting space weathering processes and remote-sensing data for the Moon and other airless bodies.

\end{abstract}




\section{Introduction}

The surfaces of airless bodies throughout the solar system, such as the Moon and asteroids, are continuously modified by a complex suite of processes collectively known as space weathering \citep{Keller1997, Pieters2016, Noble2001}. During constant exposure to the interplanetary environment, solar wind irradiation and micrometeoroid impacts fundamentally alter the physical, chemical, and spectral properties of the surface regolith \citep{Pieters2016, Noble2001}. A well-known effect of such processes is the formation of nanophase metallic iron (npFe) particles, typically less than a few tens of nanometers in diameter. These tiny iron particles are the primary agent responsible for darkening and reddening of reflectance spectra observed on mature lunar surfaces \citep{Hapke2001}. Understanding the formation mechanisms of npFe is critical for the accurate interpretation of remote sensing data and for deciphering the geologic history of planetary surfaces \citep{Pieters2016}.

A prevailing understanding over decades is that nanophase metallic iron (npFe) forms primarily through space weathering processes including. Both solar wind irradiation \citep{kuhlman2015simulation,noguchi2011incipient} and micrometeoroid impacts \citep{markley2016nanophase} contribute to npFe production. Micrometeoroid impacts generate intense localized heating and shock pressures that can vaporize or melt small volumes of iron-bearing silicate minerals abundant in the lunar regolith. 
Several mechanisms for npFe formation by micrometeoroid impact have been suggested, such as vaporization \citep{keller1993discovery,noble2005evidence}, thermal decomposition \citep{Guo2022}, and disproportionation reactions\citep{li2022impact}.
Meanwhile, solar wind irradiation can induce sputtering and chemical reduction near the surface. These conditions create a reducing environment where divalent iron (Fe$^{2+}$) is released from the silicate lattice and subsequently condenses as metallic npFe particles.

These mechanisms assume that the
source of npFe is lunar native Fe-bearing minerals such as olivine (hereafter we call this mechanism
in-situ formation).This \textit{in-situ} formation mechanism is supported by numerous laboratory experiments that sought to simulate micrometeoroid impact effects \citep{Sasaki2001,schmidt2019nanodeformation,sorokin2025experimental,fazio2018femtosecond,fulvio2021micrometeorite}. A widely used approach employs pulsed lasers to irradiate mineral samples, simulating the intense and rapid energy deposition experienced on planetary surfaces. These experiments have successfully reproduced key features of space weathering, demonstrating the npFe formation. However, an important limitation of this method is that laser ablation only simulates the energetic effects on the target material itself—it provides the thermal input but entirely excludes any contribution from the projectile. As a result, these experiments are inherently unable to capture the overall influence of impactor composition on the npFe formation, representing a methodological gap that has largely confined interpretations to processes involving native lunar materials.

Yet these factors matter significantly for a complete understanding of micrometeoroid impact.
In particular, sample analysis by \cite{zeng2025exotic} on Chang’e-5 lunar soils directly addresses this critical gap by revealing nanophase iron particles embedded within the residue of a micrometeorite impactor on an iron-poor plagioclase crystal. This finding provides the first unequivocal mineralogical evidence for an alternative, exotic source of npFe distinct from the traditional model focused solely on lunar target materials. Rather than npFe formation being confined to the reduction of native iron-bearing minerals, this study demonstrated that metallic iron could be delivered directly by the impacting micrometeorites themselves, broadening our understanding of space weathering pathways. Beyond identifying this exotic source, \cite{zeng2025exotic} estimated that such impactor-delivered npFe could contribute on the order of $\sim 5 \times 10^{6}$ tons per million years on the lunar surface, underscoring that micrometeorite-borne iron is a major, previously overlooked component of the lunar regolith’s evolution. 

Despite growing evidence from lunar sample analyses confirming multiple pathways for nanophase iron (npFe$^0$) formation \citep{zeng2025exotic,shen2024separate,Guo2022,xu2023space,lin2025differences}, critical uncertainties remain. What is the efficiency of exotic iron delivery during hypervelocity impacts? How much iron from the impactor remains on the surface as npFe versus how much is vaporized or lost? And what controls the balance between exotic iron delivery and iron reduction from native lunar minerals? Resolving these uncertainties will greatly enhance our ability to interpret the space weathering history of the Moon and other airless bodies, improving models of surface evolution and aiding future exploration efforts. Yet, the extreme temperatures, pressures, and ultrafast timescales involved challenge experimental replication and direct observation. Large-scale molecular dynamics (MD) simulations provide a unique, full-atomistic framework that can directly track the fate and physical-chemical processes of both impactor and substrate materials throughout the micrometeoroid impact event \citep{huang2021molecular,huang2025micrometeoroid,shoji2025reactive,georgiou2025effect}. Here, we use MD to conduct the first atomistic-level comparison of two npFe formation pathways: (1) exotic iron delivery from an iron-rich impactor onto an iron-poor substrate and (2) iron reduction within an iron-rich substrate impacted by an iron-poor projectile. {\color{black} In this work, we refer to ``exotic npFe" as npFe that forms from delivered iron during an impact process and thus does not mean a direct delivery of npFe included in a projectile.} By isolating these scenarios, we aim to elucidate their distinct mechanisms, quantify their relative efficiencies, and identify characteristic signatures, thereby advancing a comprehensive understanding of npFe formation and the broader space weathering process on the Moon and other airless planetary surfaces.

\section{Methods}

\subsection{ReaxFF MD simuation}
We used Reactive Force Field (ReaxFF) molecular dynamics \citep{van2001reaxff} to investigate the chemical processes active during hypervelocity impact. Unlike classical, non-reactive force fields that rely on fixed harmonic bonds, ReaxFF employs a bond-order potential in which bond orders—and thus bond energies—are recalculated at every timestep from the instantaneous interatomic distances.This enables fully dynamic simulation of bond formation, bond breaking, and short-lived reactive intermediates that are central to impact-driven space-weathering chemistry. ReaxFF also includes a geometry-dependent charge equilibration (QEq) scheme that continuously updates partial charges based on the evolving local environment, allowing accurate treatment of charge transfer and redox reactions during the micrometeoroid impact process. This capability is essential for capturing pathways such as the reduction of $\mathrm{Fe}^{2+}$ to $\mathrm{Fe}^{0}$ and the disproportionation reaction $3\mathrm{Fe}^{2+}\rightarrow\mathrm{Fe}^{0}+2\mathrm{Fe}^{3+}$. The $\mathrm{Fe/Si/O}$ ReaxFF parameterization adopted in this study was previously validated against the structural and thermodynamic properties of relevant silicate phases, ensuring its applicability to lunar regolith analogs \citep{Aryanpour2010,Buehler2007,huang2025micrometeoroid}.

\subsection{Fe-Fe Cluster Analysis}

To quantify the emergence and growth of atomic aggregates, we applied a cluster-identification method inspired by the Density-Based Spatial Clustering of Applications with Noise (DBSCAN) algorithm \citep{schubert2017dbscan}. The procedure was modified to incorporate both interatomic bonding criteria and spatial proximity. In this framework, two atoms were assigned to the same cluster if they lay within phase-appropriate bonding cutoffs (e.g., 2.400~\AA{} for Fe--Fe, 2.112~\AA{} for Fe--O, and 2.172~\AA{} for Si--O). To exclude short-lived encounters and emphasize physically meaningful structures, we also required a minimum cluster size of two atoms, analogous to the \texttt{minPts} parameter in DBSCAN. This approach enabled us to track the formation of early-stage nanophase metallic iron (npFe$^{0}$), iron oxide aggregates (e.g., Fe$_x$O$_y$), and small silicate fragments throughout the simulation. To calculate the velocity distribution function (VDF), we recorded the instantaneous velocity vectors of all individual Fe atoms within the ejecta at the end of the simulation. These velocities were then binned to construct a probability density function, which was used to estimate the fraction of Fe that escapes versus the fraction that is retained.


\subsection{Simulation Setup}

\begin{figure}[h!]
    \centering
    \begin{subfigure}[b]{0.49\textwidth}
        \includegraphics[width=\textwidth]{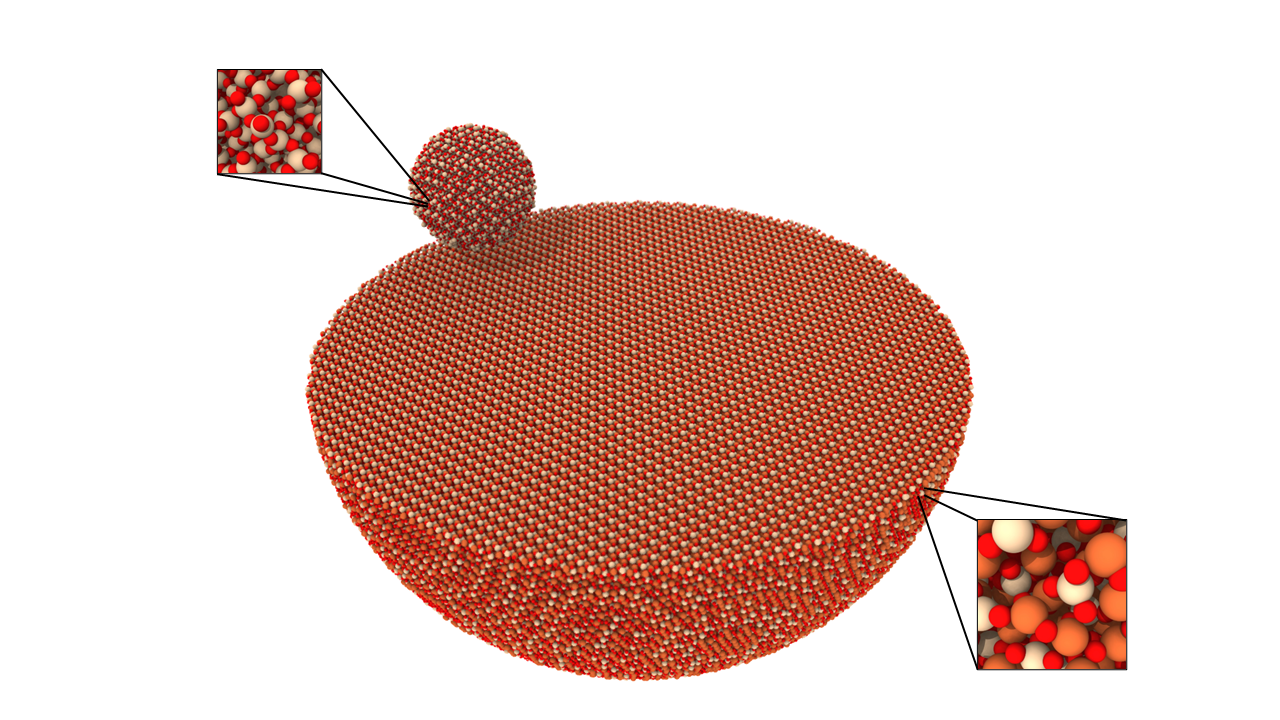}
        \caption{ }
        \label{fig:SiO2_Fe2SiO4}
    \end{subfigure}
        \hfill
    \begin{subfigure}[b]{0.49\textwidth}
        \includegraphics[width=\textwidth]{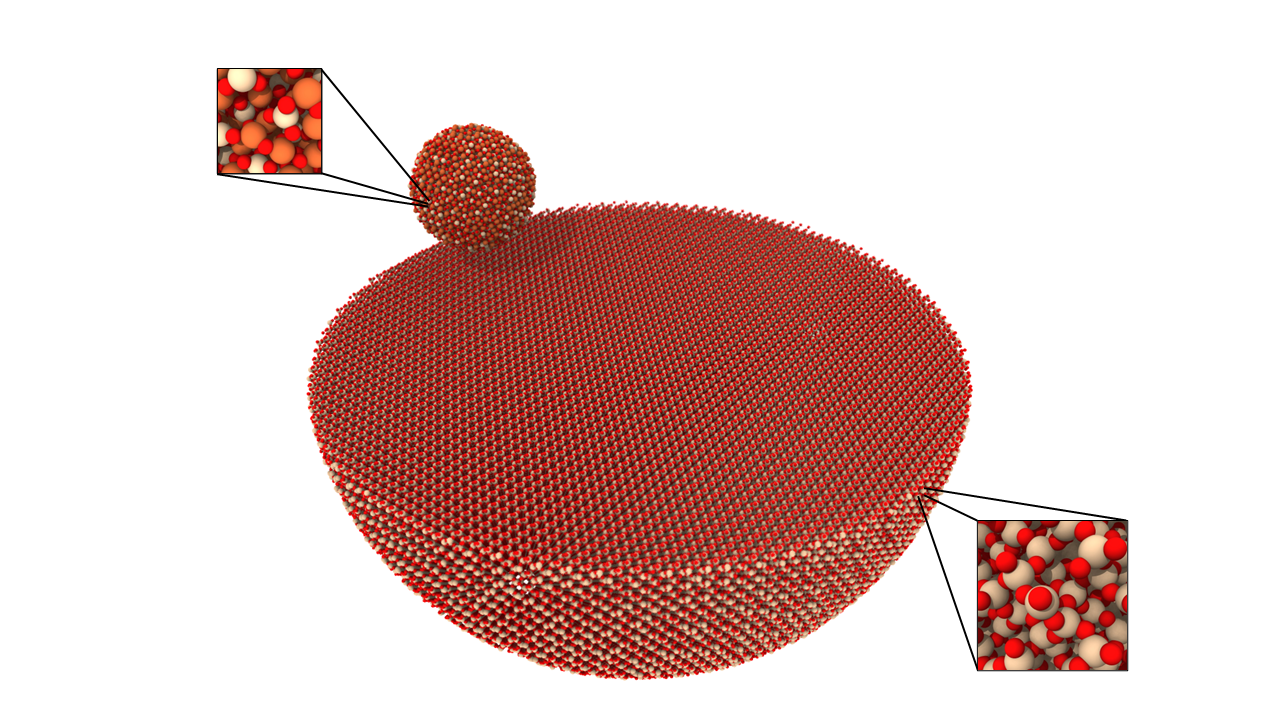}
        \caption{}
        \label{fig:Fe2SiO4_SiO2}
    \end{subfigure}
    \caption{Snapshots of the micrometeoroid impact simulations: (a) SiO$_2$ impacting Fe$_2$SiO$_4$ (Case I), (b) Fe$_2$SiO$_4$ impacting SiO$_2$ (Case II). Red particles represent oxygen, orange particles represent iron, and tan particles represent silicon. 
    }\

    \label{fig:impact_cases}
\end{figure}

To investigate the distinct pathways of nanophase iron (npFe) formation on lunar surfaces, we performed two reactive molecular dynamics (MD) simulations representing the dominant micrometeoroid-driven mechanisms summarized in Table~\ref{tab:impact_cases}  using Large-scale Atomic/Molecular Massively Parallel Simulator (LAMMPS) \cite{thompson2022lammps}. Case I corresponds to “\textit{in-situ} Fe formation,” where a SiO$_2$ micrometeoroid impacts an Fe-rich olivine surface, enabling Fe to be generated locally through impact-induced reduction and melt mixing. Case II represents “\textit{exotic} Fe delivery,” where an Fe$_2$SiO$_4$ micrometeoroid supplies Fe directly to an Fe-poor SiO$_2$ target. The initial configurations for both scenarios are illustrated in Figure~\ref{fig:impact_cases}. These paired designs differ only in the origin of Fe—produced internally versus delivered by the projectile—allowing a clean comparison of how Fe availability influences npFe nucleation. The targets and impactors were placed within a simulation domain employing periodic boundaries in the lateral $x$- and $y$-directions and a non-periodic $z$-direction that includes a vacuum region to allow ejecta to expand freely. 
To avoid the artificial shock reflections and {overheating by boundary conditions} that commonly arise in planar slab geometries with thermostat base layers, we adopted a large half-hemisphere target that can well capture the impact physics. 
The curvature and volume of this configuration allow shock energy and impact-generated heat to dissipate naturally, producing a more realistic representation of regolith response under hypervelocity impact conditions. 

Each simulation followed a three-stage protocol to ensure a stable initial state and a physically accurate impact event. The target substrate was first subjected to energy minimization to remove any initial high-energy configurations. Subsequently, the entire system was equilibrated within the canonical ensemble (NVT) which keeps the total number of atoms (N), volume of the system (V) constant and the target temperature (T) at 100 K for 10 picoseconds to achieve thermal equilibrium. Following equilibration, the bottom of the  spherical impactor was placed 6 Å above the target surface. It was assigned an initial velocity of 12 km/s, a value representative of typical micrometeoroid velocities on the Moon \citep{cintala1992impact} and with an incident angle of 45$^\circ$ towards the surface. The simulation was then run with a very small timestep of 0.1 fs to accurately integrate the equations of motion during the highly dynamic collision with microcanonical ensemble (NVE) which keeps the total number of atoms (N), volume of the system (V) and total energy (E) constant. 
The total simulation time for the impact and subsequent relaxation was extended for up to 30 ps to allow for thermal and structural evolution of the impact site until there is no new bond breaking and formation which indicates the system is stable.
Atomic trajectories, including positions, velocities, and partial charges for every atom, were saved at frequent intervals (every 100 fs) for post-processing and analysis.

\begin{table}[!h]
\centering
\caption{Summary of micrometeoroid impact simulation cases}
\label{tab:impact_cases}
\begin{tblr}{
  column{even} = {c},
  column{1} = {c},
  column{3} = {c},
  column{5} = {c},
  hline{1-2,4} = {-}{},
}
\textbf{Case} & \textbf{Projectile} & \textbf{Diameter (nm)} & \textbf{Velocity (km/s)} & \textbf{Target}  & \textbf{\textbf{Diameter (nm)}} & \textbf{\# of Atoms} \\
I             & SiO$_2$             & 6                       & 12.00    & Fe$_2$SiO$_4$                        & 32                    &         664,096   \\     
II             & Fe$_2$SiO$_4$       & 6                       & 12.00     & SiO$_2$                           & 32                         &      507,739                   \\     
\end{tblr}
\end{table}



\section{Results and Discussions}

\subsection{Fe Delivery Efficiency and Surface Retention}

In the exotic Fe delivery case, where an Fe-rich micrometeoroid impacts an Fe-poor SiO$_2$ target, we begin by tracking Fe atoms originating from the impactor. These atoms are identified by their initial positions above the impactor–target interface in the first simulation frame. By comparing their initial and final vertical positions over the course of the simulation, we distinguish Fe atoms that remain within the near-surface region from those displaced upward into the ejecta plume.
Our analysis reveals a high exotic Fe retention efficiency of approximately 91\%, comprising 68\% of material remaining in the target and 23\% being redeposited. This retention behavior for Fe is broadly consistent with that of Si ($\sim$ 72\%) and O ($\sim$ 65\%), suggesting that these elements undergo a similar mechanical retention process upon impact. Furthermore, as illustrated in Figure 3, a significant fraction of the exotic Fe—approximately 20\%—penetrates deeper than 2 nm into the target substrate.
Such high retention efficiency is central to evaluating the role of exotic Fe in micrometeoroid-driven space weathering, as only retained Fe can participate in melt chemistry, reduction processes, and the earliest stages of npFe formation.

\begin{figure}[htbp]
    \centering
    \includegraphics[width=1.0\textwidth]{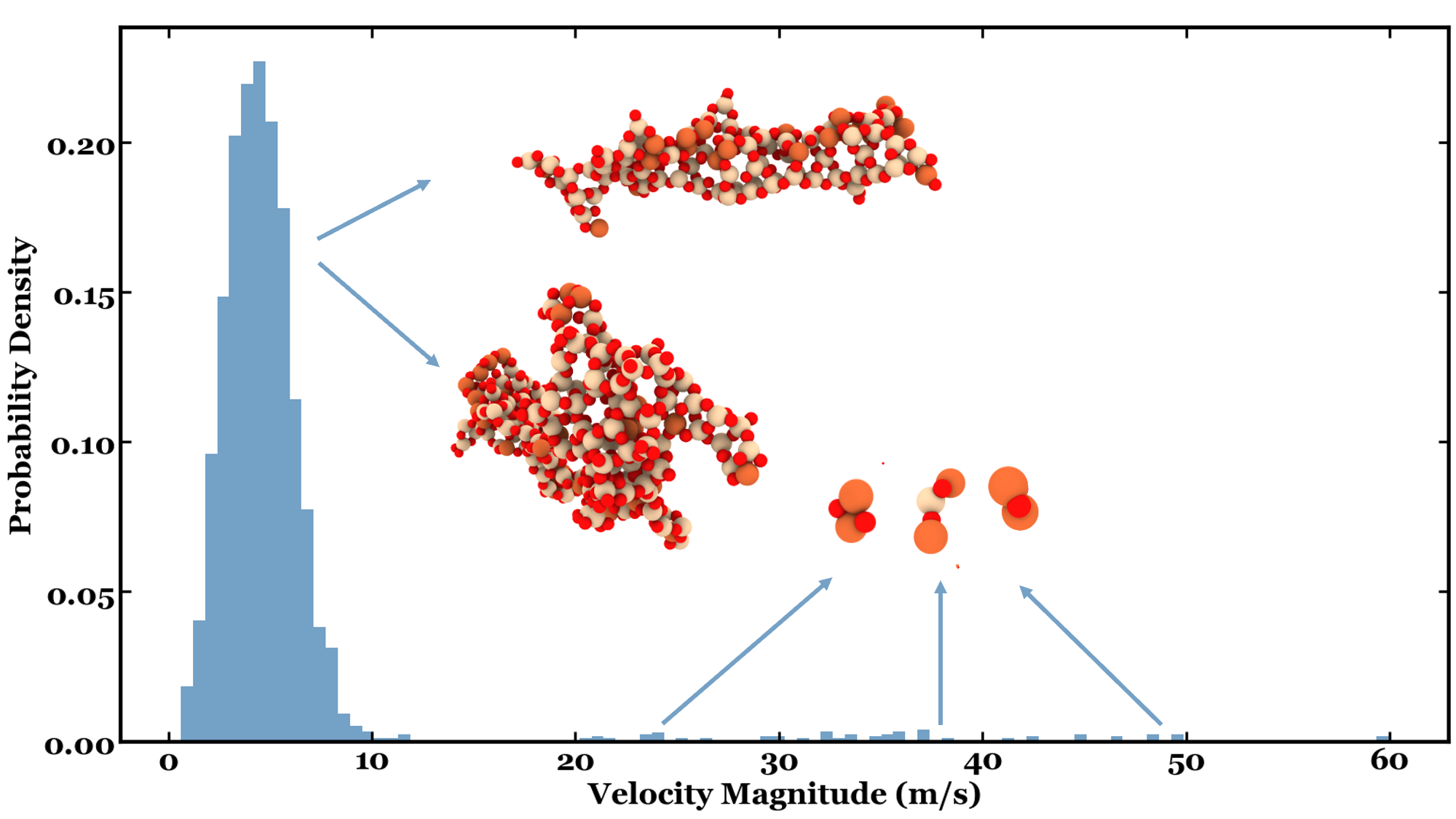}
    \caption{
    Velocity distribution function (VDF) of Fe atoms classified as ejected (above \( z=220\,\text{\AA} \)) in the exotic Fe delivery case. A pronounced Maxwellian-like peak between 0 and 10 m/s corresponds to Fe atoms bound within two large clusters, snapshots of which are shown in the inset. The sporadic velocity tail from 20 to 60 m/s represents individual atomic and molecular fragments such as Fe$_2$O$_2$, Fe$_2$SiO$_2$, and Fe$_2$O generated during the impact. This distribution suggests that most ejected Fe is locally confined and likely to re-deposit, while a smaller fraction forms transient high-velocity species. Further discussion of these molecular species is provided in \cite{huang2025micrometeoroid}.
    }
    \label{fig:exotic_fe_vdf}
\end{figure}

To further characterize the Fe atoms classified as ejected, we examined their velocity distribution, shown in Figure~\ref{fig:exotic_fe_vdf}. The distribution exhibits a pronounced low-velocity peak associated with a few coherent Fe-rich clusters. Their velocities fall well below typical lunar escape speeds (2.38 km/s), indicating that these clusters would remain gravitationally bound and undergo ballistic redeposition near the impact site. In addition to the dominant low-velocity peak, the VDF exhibits sporadic tails extending from approximately 20 m/s up to 60 m/s. These higher-velocity components correspond to individual atomic and molecular species generated during the micrometeoroid impact event, including species such as Fe\(_2\)O\(_2\), Fe\(_2\)SiO\(_2\), and Fe\(_2\)O.  The formation and distribution of these impact products reflect the complex chemistry and vaporization occurring under extreme conditions. A detailed discussion of these molecular fragments and their implications for space weathering chemistry is provided in our recent analysis of redox chemistry during impact \cite{huang2025micrometeoroid}. 
However, the majority of the ejecta found in Case II is composed of silicon oxide, because the impact predominantly ejects atoms from the target surface instead of from the impactor.
Previous pulsed laser irradiation experiments simulating micrometeorite impacts have shown that vaporization of an Fe-rich impactor can efficiently deliver iron to grain surfaces, leading to the formation of surface rims and nanophase Fe on plagioclase grains \cite{wu2017impact}. Our results support the idea that impactor-derived Fe can be transported to and incorporated into surface rims thus contributing to space weathering.
Overall, the major conclusion remains that a significant fraction of ejected Fe retains low velocities compatible with local re-deposition, implying that exotic Fe delivery primarily modifies regolith at a localized scale rather than contributing extensively to permanent iron loss.


The predominance of low-velocity ejecta indicates that many of these Fe-bearing species, even when displaced upward during the impact, 
stay within 50 meters near the impact site based on Eq. (4) in \cite{huang2021molecular}.
Their low kinetic energies make them susceptible to re-interaction with the surface, short-range collisions with other slow-moving atoms, and eventual settling within the crater. In many cases, these particles remain locally bonded within small aggregates, suggesting that “ejection” does not lead to broad spatial redistribution but instead confines Fe to a limited region surrounding the impact site. This behavior implies that exotic Fe delivered by micrometeoroids is retained not only in bulk but also in a spatially localized manner, limiting its dispersion across the regolith and preserving signatures of its origin. Such localized confinement offers insight into the fate of other exotic materials introduced by micrometeoroid bombardment and provides a foundation for future analyses of impact-driven material delivery on airless bodies.

The low-velocity nature of these atoms also raises important questions about their potential re-interaction with the surface or with other atoms, possibly influencing npFe clustering and regolith chemical heterogeneity over longer timescales. It further underscores the complexity of impactor material redistribution, where “escape” from the initial surface zone does not necessarily imply removal from the regolith system. These observations reinforce that impactor-delivered Fe can remain spatially and kinetically constrained, shaping the microscale environment of npFe nucleation. Future simulations could explore whether such low-velocity ejecta contribute significantly to lateral mixing or to the formation of impact melt deposits enriched in exotic iron.

\subsection{Post-impact Atomic Configurations and Fe-Fe bond distribution}

\begin{figure}[htbp]
    \centering
    \includegraphics[width=0.75\textwidth]{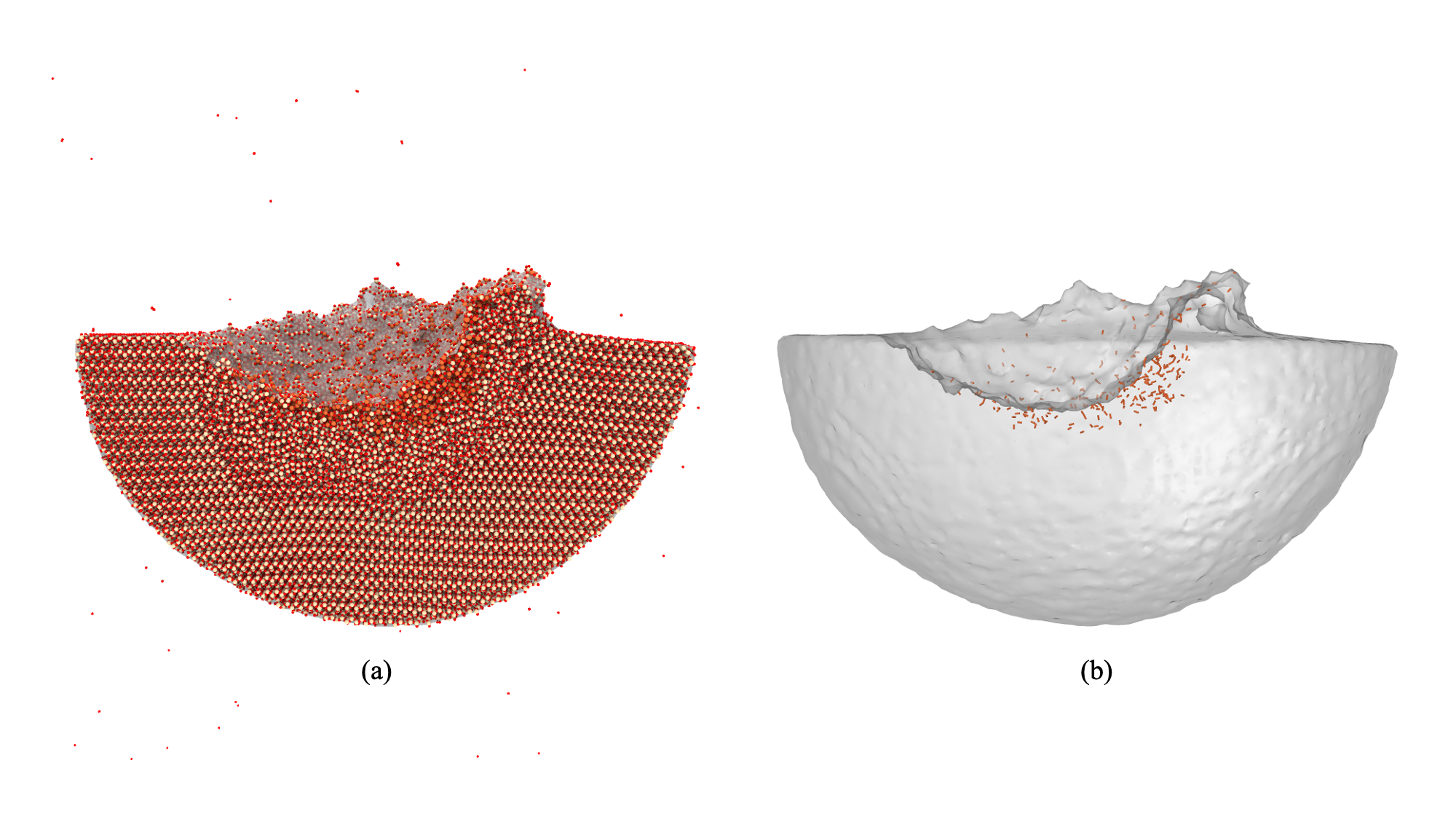}
    \caption{
    Post-impact atomic configuration for the exotic Fe delivery case. \textbf{(a)} Full atom snapshot overlaid with a surface mesh outlining the crater morphology. \textbf{(b)} Surface mesh with atoms removed, showing Fe–Fe bonds constructed using a 2.48~\AA{} cutoff threshold. Fe atoms and Fe–Fe bonds are concentrated primarily on the right side of the crater, corresponding to the downstream direction of the left-to-right impact, illustrating directional aggregation of exotic iron.
    }
    \label{fig:exotic_bond}
\end{figure}

In the exotic Fe delivery case (Figure~\ref{fig:exotic_bond}), the post-impact configuration shows a well-defined crater morphology accompanied by a concentrated accumulation of Fe atoms near the point of impact. Visualization of Fe--Fe interactions using a 2.48~\AA\ cutoff—chosen to reflect the nearest-neighbor distance in metallic bcc Fe—reveals a clear downstream-biased aggregation aligned with the projectile’s left-to-right trajectory. This spatial pattern arises because Fe delivered by the impactor is transported along the direction of peak shock propagation and melt advection, then rapidly quenched before thermal diffusion can homogenize the material. The localized clustering of Fe atoms in this region likely reflects rapid cooling and quenching conditions that stabilize npFe particles before they can diffuse away or vaporize. 
We found a maximum 4 Fe atoms in a cluster for both the exotic and in-situ impact cases.
The resulting metallic clusters therefore retain the imprint of the projectile’s motion, producing a spatially localized, trajectory-aligned Fe--Fe bonding network. 

\begin{figure}[htbp]
    \centering
    \includegraphics[width=0.75\textwidth]{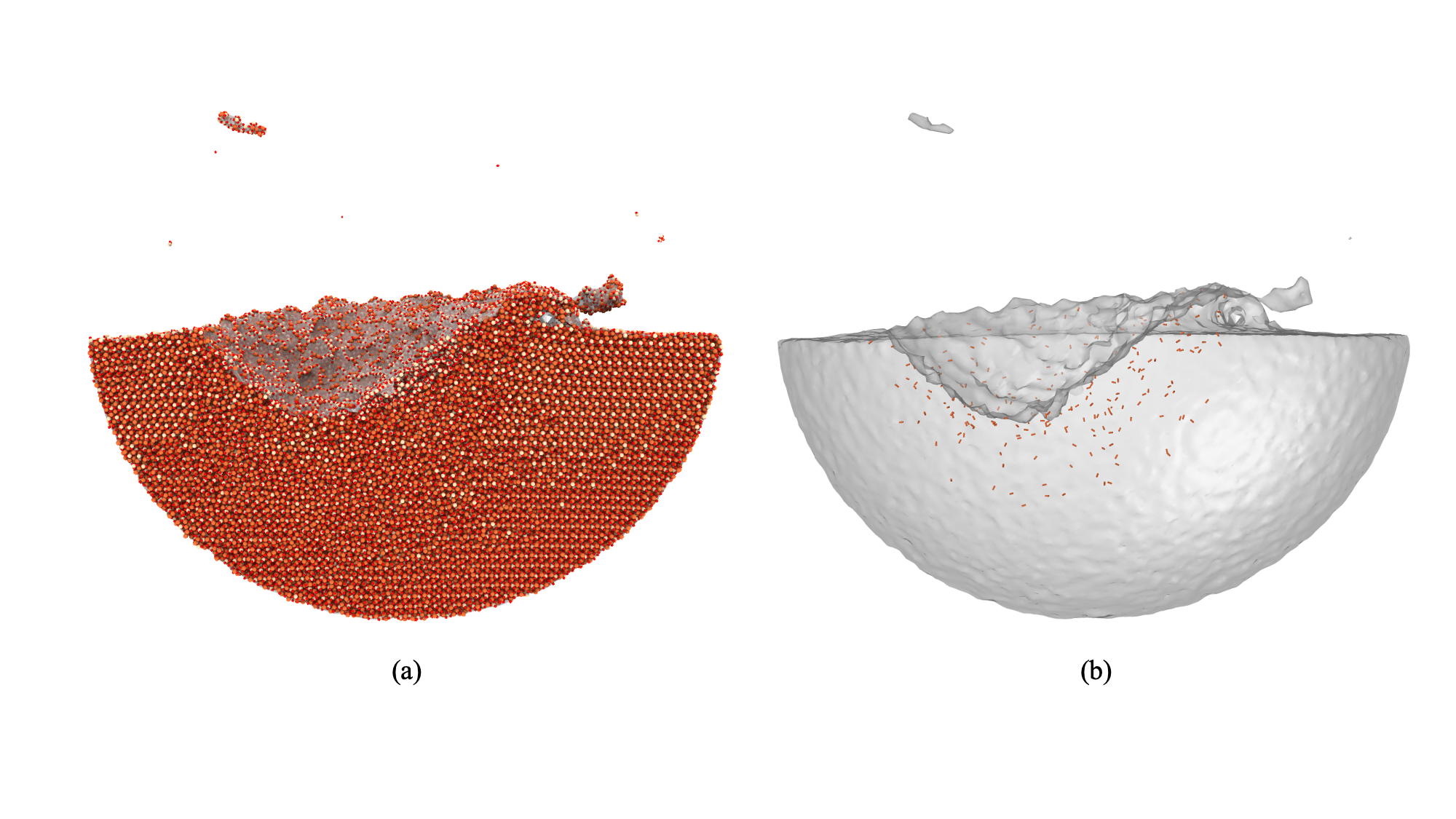}
    \caption{
    Post-impact atomic configuration for the in-situ Fe formation case. Fe atoms and Fe–Fe bonds are more diffusely distributed without clear clustering near the impact site. The spatial distribution is approximately radially symmetric, indicating npFe formation driven primarily by local thermal conditions radiating outward from the impact location.
    }
    \label{fig:insitu_bond}
\end{figure}

In contrast, the in-situ formation case (Figure~\ref{fig:insitu_bond}) exhibits a markedly different post-impact arrangement. Here, Fe originates from native iron-bearing silicates within the target, and its mobilization is governed by shock heating and subsequent thermal reduction rather than bulk transport of Fe-rich material. As heat dissipates radially from the impact center, reduction of Fe$^{2+}$ occurs in all directions, producing Fe atoms and Fe--Fe bonds that are distributed more diffusely and  uniformly. The distribution of npFe mirrors the isotropic propagation of the thermal front, consistent with npFe formation driven primarily by local high-temperature chemistry. The absence of directional bias in Fe cluster formation also suggests that npFe particles formed in this manner may be less spatially constrained and more broadly dispersed within the regolith.


These contrasting spatial signatures arise fundamentally from differences in the initial Fe source distribution. In-situ npFe forms wherever native Fe-bearing minerals experience sufficient heating to undergo reduction, leading naturally to a radially symmetric pattern. Exotic Fe, by contrast, is introduced as a concentrated reservoir carried by the projectile and advected along its trajectory, resulting in quenched metallic clusters that preserve the direction of impact-induced transport. Because these mechanisms operate under distinct spatial and thermochemical constraints, the resulting Fe--Fe  spatial distribution pattern provide a natural diagnostic of whether npFe originates from native Fe or from impactor-delivered material.

Several recent sample studies exhibit features that are suggestive of the spatial patterns predicted here. For example, Figure 3 in \citet{zeng2025exotic} shows Fe clustering with only limited apparent asymmetry, likely reflecting a section cut perpendicular to the impact direction rather than the intrinsic three-dimensional distribution. Although not capturing the full geometry, this case illustrates how sectioning orientation can obscure directional signals and highlights the need to consider spatial context when assessing npFe origins. More clearly, Figure 5b in \citet{gu2025submicron} reveals pronounced trajectory-aligned Fe-rich domains that resemble the momentum-imprinted aggregation seen in our exotic delivery simulations, and Figure 2 in \citet{li2022impact} similarly reports asymmetric clustering within localized impact features that may include exotic contributions. Collectively, these observations indicate that signatures consistent with impactor-derived Fe are already present within the lunar sample archive, even if they have not yet been systematically distinguished from in-situ processes. Based on our results, we predict that exotic Fe formation is likely to be widespread—if not ubiquitous—among materials returned by the Apollo and Chang’e missions, awaiting detailed spatial analysis to reveal its full extent.

These mechanistic distinctions provide a powerful framework for interpreting Chang’e-5 samples and future lunar returns. The contrasting spatial patterns mean that exotic and in-situ npFe can be differentiated not through surrounding mineral stoichiometry, as suggested by \citet{zeng2025exotic}, nor through isotopic or compositional measurements that can be ambiguous or difficult to acquire, but through their geometric distribution. Spatial arrangement offers a far more direct and experimentally accessible diagnostic: the trajectory-imprinted, momentum-aligned npFe expected from exotic delivery contrasts sharply with the radially symmetric, thermally driven npFe produced in situ. 
Because these spatial signatures persist even in Fe-rich regolith where both mechanisms may operate simultaneously, they provide a robust means to disentangle overlapping formation pathways. 
Such spatial diagnostics could complement the fundamental mineralogical and crystallographic analysis central to sample studies. While interpreting these patterns requires careful consideration of FIB sectioning orientation and potential redistribution from solar wind irradiation, this approach provides a valuable supporting tool to enhance the scientific return of future sample analysis.

\section{Conclusion}

Our simulations provide the first atomistic comparison of exotic versus in-situ npFe formation during micrometeoroid bombardment and reveal that these two pathways leave distinct and interpretable nanoscale signatures. Exotic Fe delivered by olivine impactors is efficiently retained and quenched into localized, trajectory-aligned clusters that preserve the memory of projectile transport, whereas in-situ npFe emerges diffusely and radially from shock-driven reduction of native Fe-bearing minerals. These contrasting spatial patterns offer a practical diagnostic that complements, and in many cases is more direct than, approaches based on surrounding mineral stoichiometry \citep{zeng2025exotic} or isotopic composition. Because spatial geometry can be assessed rapidly with high-resolution TEM or STEM-EELS, this metric provides an accessible means to distinguish the origins of npFe in returned samples. These findings carry broader implications for long-standing problems in lunar space weathering, including the interpretation of spectral maturation trends \cite{lucey2014global}, and the variability of space-weathered rims across different lithologies \cite{gu2025submicron}.  Understanding the balance between these processes will also inform how micrometeoroid impacts contribute to impact gardening and mixing of regolith materials \cite{costello2018mixing,costello2020impact}, and the redistribution of metal-rich grains across the surface.

In addition to the two pathways directly captured in our simulations, micrometeoroids may also deliver pre-formed npFe embedded within their own space-weathered rims. Outstanding questions remain regarding the survivability of such particles: whether they remain intact upon impact, whether their bcc Fe structure is preserved, and how deeply they may implant into the regolith. Resolving these questions will require combined experimental, atomistic, and sample-based investigations and represents an important direction for future work. Subsequent studies will investigate the effects of impact velocity and angle on npFe$^0$ distribution.
Taken together, these results provide a coherent framework for interpreting both Apollo and Chang’e samples and for guiding future Artemis-era analyses. By linking npFe spatial signatures directly to their formation origin, this approach enables targeted identification of exotic contributions and offers a practical pathway for disentangling mixed formation histories in lunar soils.

\begin{acknowledgments}

This research was supported in part through research cyber infrastructure resources and services provided by the Partnership for an Advanced Computing Environment (PACE) at the Georgia Institute of Technology, Atlanta, Georgia, USA. Z.H. and M.H. are supported by SSERVI-CLEVER (NNH22ZDA020C/80NSSC23M022). M.H. acknowledges support by VIPER (80NSSC24K0682) and SSERVI/RASSLE (80NSSC24M0016).

\end{acknowledgments}

\begin{contribution}
ZH: Conceptualization, Data curation, Formal analysis, Investigation, Methodology,  Software, Validation, Visualization, Writing – original draft, Writing – review \& editing.
M.H.: Investigation, Project administration, Funding acquisition Supervision, Writing – review \& editing.



\end{contribution}

%

\software{scipy \cite{virtanen2020scipy}, LAMMPS \cite{thompson2022lammps}, OVITO \cite{stukowski2009visualization}
          }






\bibliography{spaceweatheringnpFe}{}
\bibliographystyle{aasjournalv7}



\end{document}